\documentclass[pra,a4]{revtex4}

\begin{document}

\markboth{Astrid Lambrecht and Serge Reynaud} {Casimir Effect :
Theory and Experiments}

% -- My Definitions --
\def\Cas{\mathrm{Cas}}
\def\P{\mathrm{P}}    %plasma
\def\dd{\mathrm{d}}
\def\max{\mathrm{max}}
\def\min{\mathrm{min}}
\def\patch{\mathrm{patch}}
\def\rms{\mathrm{rms}}
\def\calF{\mathcal{F}}
%%%%%%%%%%%%%%%%%%%%%%%%%%%%%%%%%%%%%%%%%%%%%%%%%%%%%%%%%%%%%%%%%%%%

\title{\textbf{CASIMIR EFFECT : THEORY AND EXPERIMENTS}  }

\author{ASTRID LAMBRECHT and SERGE REYNAUD}

\address{Laboratoire Kastler Brossel, CNRS, ENS, UPMC \\
Campus Jussieu, F75252 PARIS (France)}

\begin{abstract}
The Casimir effect is a crucial prediction of Quantum Field Theory
which has fascinating connections with open questions in fundamental
physics. The ideal formula written by Casimir does not describe real
experiments and it has to be generalized by taking into account the
effects of imperfect reflection, thermal fluctuations, geometry as
well as the corrections coming from surface physics. We discuss
these developments in Casimir physics and give the current status in
the comparison between theory and experiment after years of
improvements in measurements as well as theory.
\end{abstract}
\maketitle

\section{Introduction}

The Casimir effect \cite{Casimir48} deserves careful attention as a
crucial prediction of Quantum Field Theory
\cite{Milonni94,LamoreauxAJP99,BarreraNJP06,CasimirPhysics}. It also
has fascinating interfaces with other open questions in fundamental
physics.

It is connected with the puzzles of gravitational physics through
the problem of vacuum energy \cite{GenetDark02,JaekelMass08} as well
as with the principle of relativity of motion through the dynamical
Casimir-like effects \cite{JaekelRPP97,LambrechtJOB05}. Effects
beyond the Proximity Force Approximation also make apparent the rich
interplay of vacuum and geometry \cite{BalianPoincare03}.

Casimir physics also plays an important role in the tests of gravity
at sub-millimeter ranges \cite{Fischbach98,AdelbergerARNPS03}. For
scales of the order of the micrometer, gravity tests are performed
by comparing Casimir force measurements with theory
\cite{LambrechtPoincare03,DeccaEPJC07,LambrechtCasimir11}. Other
constraints can be obtained with atomic or nuclear force
measurements \cite{AntoniadisCRAS11}.

Finally, the Casimir force and closely related Van der Waals force
have strong connections with various active domains and interfaces
of physics, such as atomic and molecular physics, condensed matter
and surface physics, chemical and biological physics, micro- and
nano-technology~\cite{Parsegian06}.

\section{The puzzle of vacuum energy}

The first quantum law was written by Planck in 1900 to explain the
properties of the black body radiation~\cite{Planck00}. In modern
terms, it gives the mean energy per electromagnetic mode as the
product $\overline{n}\hbar\omega$ of the photon energy $\hbar
\omega$ by the mean number of photons per mode
$\overline{n}=\left(\exp\frac{\hbar\omega}{k_\mathrm{B}T}-1\right)^{-1}$.

In 1911, Planck \cite{Planck12} wrote a new expression for the mean
energy per mode $\left(\overline{n}+\frac12\right)\hbar\omega$ which
contained a zero-point energy $\frac12\hbar\omega$ besides the black
body energy. In contrast to the latter, the zero-point fluctuations
were still present at zero temperature. The arguments thus used by
Planck cannot be considered as consistent today. The first known
argument still acceptable today was proposed by Einstein and
Stern~\cite{Einstein13} in 1913~: the modified Planck law reproduces
the classical limit at high temperatures
$\left(\overline{n}+\frac12\right)\hbar\omega= k_{{\rm B}}T+O\left(
\frac{1}{T}\right)$.

In 1916, Nernst discussed zero-point fluctuations for the
electromagnetic field and discovered that their energy constituted a
challenge for gravitation theory~\cite{Nernst16}. When summing up
the zero-point energies over all field modes, a finite energy
density is obtained for the first Planck law (this is the solution
of the `ultraviolet catastrophe') but an infinite value is produced
from the second law. When introducing a high frequency cutoff, the
calculated energy density remains finite but it is still much larger
than the mean energy observed in the world around us through
gravitational phenomena~\cite{WeinbergRMP89}.

This major problem has led famous physicists to deny the reality of
vacuum fluctuations. In particular, Pauli stated in his textbook on
Wave Mechanics~\cite{Pauli33} : \emph{At this point it should be
noted that it is more consistent here, in contrast to the material
oscillator, not to introduce a zero-point energy of
$\frac12\hbar\omega$ per degree of freedom. For, on the one hand,
the latter would give rise to an infinitely large energy per unit
volume due to the infinite number of degrees of freedom, on the
other hand, it would be principally unobservable since nor can it be
emitted, absorbed or scattered and hence, cannot be contained within
walls and, as is evident from experience, neither does it produce
any gravitational field.}

A part of these statements is simply unescapable~: it is just a
matter of evidence that the mean value of vacuum energy does not
contribute to gravitation as an ordinary energy. But it is certainly
not possible to uphold that vacuum fluctuations have no observable
effects. Certainly, vacuum fluctuations are \emph{scattered} by
matter, as shown by their numerous effects in
atomic~\cite{CohenTannoudji92} and subatomic~\cite{Itzykson85}
physics. And the Casimir effect is nothing but the evidence of
vacuum fluctuations making their existence manifest when being
\emph{contained within walls}.

\section{The Casimir force}

Casimir calculated the force in an idealized case, with perfectly
smooth, flat and parallel plates in the limit of zero temperature
and perfect reflection. The expressions for the energy $E_\Cas= -
\hbar c \pi^2 A/720 L^3$ and force $F_\Cas=-\dd E_\Cas/\dd L$ thus
reveal a universal behavior resulting from the confinement of vacuum
fluctuations ($L$ is the distance, $A$ the area, $c$ the speed of
light and $\hbar$ the Planck constant).

For the metallic mirrors used in experiments, the force depends on
the optical properties of the mirror described by a dielectric
function \cite{LambrechtEPJ00}. The dielectric function is a sum of
contributions corresponding to interband transitions and conduction
electrons. The latter contribution is directly related to the
frequency dependent conductivity of the metal $\sigma [\omega] =
\omega_\P^2/(\gamma-i\omega)$ where $\omega_\P$ is the plasma
frequency and $\gamma$ the Drude damping constant. As $\gamma$ is
much smaller than $\omega_\P$ for a good metal such as Gold, the
limiting lossless case ($\gamma \to 0$) captures a large part of the
effect of imperfect reflection. However it is not an accurate
description: a much better fit of tabulated optical data is obtained
with a non null value of $\gamma$; moreover, a dissipative Drude
model ($\gamma\neq0$) meets the well-known fact that Gold has a
finite static conductivity $\sigma_0 = \omega_\P^2/\gamma$.

Experiments are performed at room temperature so that the effect of
thermal fluctuations has to be added to that of vacuum fields
\cite{Mehra67,Brown69,SchwingerAP78,GenetPRA00}. Bostr\"{o}m and
Sernelius were the first to remark that, despite its small value,
$\gamma$ had a large effect on the force at non null temperatures
\cite{BostromPRL00}. In particular, the ratio between the
predictions calculated for the lossless and lossy cases reaches a
factor 2 at large distances. A large number of contradictory papers
has been devoted to this topic (see references in
\cite{MiltonJPA05,KlimchitskayaCP06,BrevikNJP06,IngoldPRE09}) and
the contradiction is deeply connected to the comparison between
theory and experiments discussed below. It is also worth recalling
here that derivations from microscopic models of the metallic
mirrors give, as should be expected, Casimir forces agreeing at
large distance with predictions of the dissipative Drude model
\cite{Jancovici05,Buenzli05,Bimonte09}.

Another important feature of the recent precise experiments is that
they are performed in the plane-sphere geometry. The estimation of
the force in this geometry uses the so-called \textit{Proximity
Force Approximation} (PFA) \cite{DerjaguinQR68} which amounts to
integrate over the distribution of local inter-plate distances the
pressure calculated in the geometry with two parallel planes. This
approximation can only be valid as a limit for sphere radius $R$
much larger than the separation $L$. Even in this case its accuracy
remains a question of importance for the comparison between theory
and experiments discussed in the sequel of this paper. This question
has been studied in recent papers
\cite{EmigJSM08,MaiaPRA08,CanaguierPRL09,CanaguierPRL10,CanaguierPRA10}
and the related advances are presented elsewhere in this volume
\cite{CanaguierThisVolume}.

\section{The scattering approach to the Casimir effect}

For preparing forthcoming discussions, it is worth surveying the
scattering approach which is the best tool available today for
calculating the Casimir effect \cite{LambrechtNJP06}.

The basic idea is that mirrors are described by their scattering
amplitudes. It can be simply illustrated with the model of scalar
fields propagating along the two directions on a line (1-dimensional
space; see references in \cite{JaekelRPP97}). Each mirror is
described by a scattering matrix containing reflection and
transmission amplitudes. Two mirrors form a Fabry-Perot cavity
described by a scattering matrix $S$ which can be deduced from the
two elementary matrices. The Casimir force then results from the
difference of radiation pressures exerted onto the inner and outer
sides of the mirrors by the vacuum field fluctuations
\cite{JaekelJP91}. Equivalently, the Casimir free energy can be
written as the sum of the frequency shifts of all vacuum field modes
due to the presence of the cavity \cite{JaekelJP91}.

The same discussion can be extended to the geometry of two plane and
parallel mirrors aligned along the axis $x$ and $y$, described by
specular reflection and transmission amplitudes which depend on the
frequency, transverse wavevector and polarization \cite{JaekelJP91}.
A few points have to be treated with care when extending the
derivation from 1-dimensional space to 3-dimensional space~:
evanescent waves contribute besides ordinary modes freely
propagating outside and inside the cavity; dissipation has to be
accounted for~\cite{GenetPRA03}. The properties of the evanescent
waves are described through an analytical continuation of those of
ordinary ones, using the well defined analytic behavior of the
scattering amplitudes. At the end of this derivation, this analytic
properties are also used to perform a Wick rotation from real to
imaginary frequencies. This leads to an expression for the Casimir
free energy under the form of a Matsubara formula
\cite{LambrechtCasimir11}.

This formula reproduces the Casimir ideal formula in the limits of
perfect reflection $r \rightarrow 1$ and null temperature $T
\rightarrow 0$. But it is valid and regular at thermal equilibrium
at any temperature and for any optical model of mirrors obeying
causality and high frequency transparency properties. It can thus be
used for calculating the Casimir force between arbitrary mirrors, as
soon as the reflection amplitudes are specified. These amplitudes
are commonly deduced from models of mirrors, the simplest of which
is the well-known Lifshitz model
\cite{LifshitzJETP56,DzyaloshinskiiUspekhi61}. In this model,
semi-infinite bulk mirrors are characterized by a linear and local
dielectric response function and reflection amplitudes are then
deduced from the Fresnel law. It is worth emphasizing that the
scattering formula allows to accommodate more general expressions
for the reflection amplitudes. In particular, it may be used even
when the optical response of the mirrors can no longer be described
by a local dielectric response function. The reflection amplitudes
can for example be determined from microscopic models of mirrors.
Recent attempts in this direction can be found for example in
\cite{Pitaevskii08,Geyer09,Dalvit08b,Decca09,Svetovoy08}.

\section{Casimir experiments}

We now discuss the status of Casimir experiments and their
comparison with theory. At this point, we face the difficulties that
there are persisting differences between experimental results and
theoretical predictions drawn from the best motivated models, as
well as disagreements between some recent experiments.

On one hand, there have been experiments at IUPUI and UCR for
approximately ten years, with results pointing to an unexpected
conclusion \cite{DeccaAP05,DeccaPRD07,KlimchitskayaRMP09}. In
particular, the Purdue experiment uses dynamic measurements of the
resonance frequency of a microresonator. The shift of the resonance
gives the gradient of the Casimir force in the plane-sphere
geometry, which is proportional (within PFA) to the Casimir pressure
between two planes. The typical radius of the sphere is $R=150\mu$m
and the range of distances $L=0.16-0.75\mu$m. The results appear to
fit predictions obtained from the lossless plasma model $\gamma=0$
rather than those corresponding to the expected dissipative Drude
model $\gamma\neq0$ (see Fig.1 in~\cite{DeccaPRD07}), in
contradiction with the fact that Gold has a finite static
conductivity. IUPUI and UCR experiments are performed at distances
where the thermal contribution as well as the effect of $\gamma$ are
not large, so that the estimation of accuracy is a critical issue.

On the other hand, a new experiment in Yale \cite{SushkovNatPh11}
uses a much larger sphere $R=156$mm, which allows for measurements
at larger distances $L=0.7-7\mu$m. The thermal contribution is large
there and the difference between the predictions at $\gamma=0$ and
$\gamma\neq0$ is significant. The results of this experiment see the
thermal effect and they fit the predictions drawn from the
dissipative Drude model, after the contribution of the electrostatic
patch effect has been subtracted
\cite{SushkovNatPh11,LamoreauxCasimir11}. Of course, these new
results have to be confirmed by further studies
\cite{MiltonNatPh11}. The main issue in this experiment is that the
pressure due to electrostatic patches is larger than that due to
Casimir effect, and that the patch distribution is not characterized
independently. This is in fact a more general problem since the
patch distribution is not measured in other experiments either (more
discussions below).

In this short discussion, we have focused our attention on a few
experiments. For completeness, we give here a list of other Casimir
measurements between metallic plates which have produced information
of interest on the topics discussed in this paper
\cite{EderthPRA00,ChanPRL01,BressiPRL02,LisantiPNAS05,SvetovoyPRB08,%
OnofrioPRA08,JourdanEPL09,deManPRA09,MasudaPRL09}.

\section{Discussion}

The conclusion at this point is that the Casimir effect, now
measured in several experiments, is not yet tested at the 1\% level,
as has been sometimes claimed. While the Yale experiment fits
predictions drawn from the dissipative Drude model, IUPUI and UCR
experiments favor theoretical predictions obtained with the lossless
plasma model. When comparing the IUPUI experimental data with the
predictions drawn from the best motivated model (the dissipative
Drude model), the pressure difference goes up to $\sim$50mPa at the
smallest distances $\sim$160nm where the pressure itself is
$\sim$1000mPa. This difference is clearly larger than the
statistical dispersion (see Fig.1 in~\cite{DeccaPRD07}).

This question is important not only for the test of the Casimir
effect, a central prediction of Quantum Field Theory, but also
because Casimir experiments are one of the main routes in the search
for hypothetical new short-range forces beyond the standard model
\cite{Fischbach98,AdelbergerARNPS03,AntoniadisCRAS11}. The
difference between IUPUI experimental data and theoretical
predictions (using the Drude model) does not look like a Yukawa
force law, but it looks like a superposition of power laws which
meet predictions of some currently considered unification models
\cite{AntoniadisCRAS11}.

This discrepancy between theory and experiment may have various
origins, in particular artefacts in the experiments or inaccuracies
in the calculations. They may also come from yet unmastered
systematic effects in the analysis of experimental data. In
particular recent publications study the effects of surface physics
on Casimir experiments. Electrostatic patches, already alluded to
above, are a probable source of such systematic effects (more
discussions below). The problem of surface roughness has also to be
studied in a thorough manner
\cite{MaiaNetoEPL05,MaiaNetoPRA05,vanZwolPRB09,BroerEPL11,vanZwolCasimir11}.

In the sequel of this paper, we discuss the effect of electrostatic
patches which is a known limitation for a large number of high
precision measurements \cite{FairbankPRL67,CampJAP91,TurchettePRA07,%
DeslauriersPRL06,RobertsonCQG06,EpsteinPRA07,PollackPRL08,%
AdelbergerPPNP09,EverittPRL11,ReasenbergCQG11}, and is in particular
for Casimir experiments \cite{SpeakePRL03,ChumakPRB04,%
KimPRA10,KimJVST10,deManJVST10,Behunin11}. The patch effect is due
to the fact that the surface of a metallic plate cannot be an
equipotential as it is constituted by micro-crystallites with
different work functions. For clean metallic surfaces studied by the
techniques of surface physics, the resulting ``voltage roughness''
is correlated to the ``topography roughness'' \cite{Gaillard06}. For
surfaces exposed to air, the situation is changed due to the
unavoidable contamination by adsorbents. The latter is known to
spread out the electrostatic patches, enlarge correlation lengths
and reduce voltage dispersions \cite{Rossi92}.

The pressure due to electrostatic patches between two planes can be
computed exactly by solving the Poisson equation \cite{SpeakePRL03}.
Its evaluation only depends on the spectra describing the
correlations of the patch voltages or, equivalently on the
associated noise spectra. In most analysis of the patch pressure
devoted up to recently to Casimir experiments
\cite{DeccaAP05,DeccaPRD07}, the spectrum has been assumed to be
flat between two cutoffs (a minimum wavevector $k_\min$ and a
maximum one $k_\max$), which is a very poor representation of the
patches on real surfaces. A ``quasi-local'' model has recently been
proposed which gives a much better motivated representation of
patches \cite{Behunin11}. The model is based on a tessellation of
the sample surface and a random assignment of the voltage on each
patch. It produces a smooth spectrum different from the
``sharp-cutoff'' model used in previous analysis
\cite{DeccaAP05,DeccaPRD07}. It is worth emphasizing that the new
quasi-local model shows close similarities with models used recently
to explain the effect of patches on heating in ion or atom traps and
cantilever damping \cite{Dubessy09,Carter11}.

The results of the calculations of patch pressure are described in
\cite{Behunin11}. When the patch effect is estimated with the
sharp-cutoff model and the same parameters as in
\cite{DeccaAP05,DeccaPRD07}, a very small contribution is obtained
which cannot explain the difference between experimental data and
theoretical predictions using the Drude model. In contrast, when the
patch effect is estimated with the quasi-local spectrum and
parameters deduced from the grain sizes as in
\cite{DeccaAP05,DeccaPRD07}, an unexpected result is obtained: the
calculated patch pressure is now larger than the residuals
(difference between experimental data and theoretical predictions
using the Drude model). This means that patches are an important
systematic effect for Casimir force measurements and that the patch
spectrum should ideally be characterized independently of this
measurement.

As the computed patch pressure is obviously model dependent, it
seems natural to try to find a model between the two cases presented
above which would reproduce as least qualitatively the residuals.
This question has been addressed in \cite{Behunin11} by varying the
parameters of the better motivated quasi-local model. It was found
that the output of the model depends mainly on two parameters,
namely the size of largest patches $\ell_\patch^\max$ and the rms
voltage dispersion $V_\rms$. A best-fit on these two parameters
produces an agreement between the residuals and the patch pressure.
The best-fit values for the parameters $\ell_\patch ^\max$ and
$V_\rms$ are quite different from those which would be obtained by
identifying patch sizes to crystallite sizes. However, as
$\ell_\patch ^\max$ is larger than the maximum grain size $\sim
300$nm, and $V_\rms$ smaller than the rms voltage $\sim 81$mV which
would be associated with the random work functions of
micro-crystallites, these values are reasonable for contaminated
surfaces \cite{Rossi92}.

These results mean that the difference between IUPUI experimental
data and theoretical predictions can be reproduced at least
qualitatively by the quasi-local model for electrostatic patches.
Let us stress at this point that this is the result of a fit, with
the parameters of the patch model not measured independently (the
same criticism holds as well for previous analysis of the data). A
better characterization of the patches is a crucial condition to
reaching firmer conclusions. The patch distributions can be measured
with the dedicated technique of Kelvin probe force microscopy (KPFM)
which is able to achieve the necessary size and voltage resolutions
\cite{Liscio08,Liscio11}. In addition, the study of cold atoms and
cold ions trapped in the vicinity of metallic surfaces
\cite{EpsteinPRA07,Dubessy09,Carter11} or the role of patch effects
in other precision measurements
\cite{AdelbergerPPNP09,EverittPRL11,ReasenbergCQG11} are other ways
for accessing information of interest for our problem.

\section*{Acknowledgments}
The authors thank R.O. Behunin, A. Canaguier-Durand, I.
Cavero-Pelaez, J. Chevrier, T. Coudreau, D.A.R. Dalvit, R.S. Decca,
T. Ebbesen, C. Genet, R. Gu\'erout, G.-L. Ingold, F. Intravaia,
M.-T. Jaekel, S.K. Lamoreaux, J. Lussange, P.A. Maia Neto, V.V.
Nesvizhevsky, P. Samori, S. Seidelin and C. Speake for contributions
or discussions related to this paper, and the ESF Research
Networking Programme CASIMIR (www.casimirnetwork. com) for providing
excellent opportunities for discussions on the Casimir effect and
related topics.

%\section*{References}

\end{document}